\documentclass[]{aa}  
\usepackage{graphicx}
\usepackage{natbib}  
\bibpunct{(}{)}{;}{a}{}{,}
\usepackage{txfonts}
\usepackage{hyperref}

\begin{document}

\title{Updated astrometry and masses of the LUH 16 brown dwarf binary}
\titlerunning{Updated astrometry and masses of LUH 16}
\author{P. F. Lazorenko\inst{1}
                \and J. Sahlmann\inst{2}}

\institute{Main Astronomical Observatory, National Academy of Sciences of the Ukraine, Zabolotnogo 27, 03680 Kyiv, Ukraine\\
                \email{laz@mao.kiev.ua}                         
                \and
                Space Telescope Science Institute, 3700 San Martin Drive, Baltimore, MD 21218, USA}

\date{Received ; accepted  }
\abstract{The nearest known binary brown dwarf \object{WISE J104915.57-531906.1AB} (LUH 16) is a well-studied benchmark for our understanding of substellar objects. Previously published astrometry  { of LUH 16 obtained with FORS2 on the Very Large Telescope} was affected by errors that limited its use in combination with other datasets, thereby hampering the determination of its accurate orbital parameters and masses.
We improve upon the calibration and analysis of the FORS2 astrometry with the help of Gaia DR2 to generate a high-precision dataset that can be combined with present and future LUH 16 astrometry. We demonstrate its use by combining it with available measurements from the Hubble Space Telescope (HST) and Gemini/GeMS and deriving updated orbital and mass parameters.
Using Gaia DR2 as astrometric reference field, we derived the absolute proper motion  and updated the absolute parallax of the binary to 501.557$\pm$0.082~mas. We refined the individual dynamical masses of LUH 16 to $33.5\pm 0.3\, M_{Jup}$  (component A) and $28.6\pm 0.3\,M_{Jup}$ (component B), which corresponds to a relative precision of $\sim$1\% and is three to four times more precise than previous estimates. { We found that these masses show a weak dependence on one datapoint extracted from a photographic plate from 1984. The exact determination of a residual mass bias, if any, will be possible when more high-precision data can be incorporated in the analysis.} 
}
\maketitle

\section{Introduction}
After its discovery by \citet{Luhman}, the binary brown dwarf WISE J104915.57-531906.1 (Luhman 16, hereafter LUH 16) located at 2 pc from the Sun has been observed extensively. The astrometric follow-up by \citet{Boffin} used observations with the FORS2 camera on the Very Large Telescope (VLT) between April and June of 2013 to refine the parallax value to $2.020\pm 0.019$ pc and to claim indications for the presence of a massive substellar companion around one of the binary components. 

Including additional { epochs  obtained in 2014 with FORS2,} \citep{2015MNRAS} derived the relative (500.23 mas) and absolute parallax $\varpi_\mathrm{abs}=500.51 \pm 0.11$~mas of this system,  obtained an upper mass limit for a potential third body of 2$M_{\rm Jup}$, and determined a mass ratio $q=0.78 \pm 0.10$  for the LUH 16 binary.  After publication of Gaia Data Release 1  \citep[DR1][]{GaiaCollaboration:2016aa, DR1_coll, Lindegren:2016aa}, we updated the FORS2 distortion correction, which led to the updated  values of $501.139$ mas and $\varpi_\mathrm{abs}=501.419 \pm 0.11$~mas for the relative  and absolute  parallax, respectively \citep{gaia}.

Using Hubble Space Telescope (HST) observations in 2014--2016,  \citet{Bedin} derived a relative $501.118 $~mas and absolute parallax $501.398 \pm 0.093$~mas for LUH 16. Because of a longer observation time span, they were able to obtain estimates of the orbital elements of the system, in particular,  $31.3\pm 7.9$ years for the orbital period and $0.463 \pm 0.064$ for eccentricity.   \citet{Bedin}  found inconsistencies between their HST astrometry and the \citet{2015MNRAS} astrometry at the level of 10--20~mas and therefore did not use FORS2 measurements for the orbit fitting. 

\citet{Garcia} performed an independent analysis of the FORS2 observations obtained in 2013--2014 and added five more epochs in 2015, using Gaia DR1 for the transformation from CCD positions to the International celestial reference frame (ICRF). They added relative astrometry from the Gemini South Multiconjugate Adaptive Optics System (GeMS,  \citealt{Ammons}), {ESO photographic plates,}  { CRIRES radial velocity,} and archival observations from the Deep Near-Infrared Survey of the Southern Sky (DENIS). The HST astrometry was not published at that time and was not used. 

Here we report new results from the combination of all available datasets, which increases the time span of high-precision astrometry from $\sim$2 years covered by the \citet{Bedin} study to 3.5 years. In comparison with our first study \citep{2015MNRAS}, we present improved astrometric calibrations of the FORS2 data. We then update the orbital parameters and dynamical masses of the binary on the basis of a combination of  FORS2, HST, GeMS, {ESO archive,} Gaia DR1, and Gaia DR2 \citep{DR2, 2018yCat} astrometry,  { and the CRIRES relative radial velocity to constrain the  inclination.}

\section{FORS2 astrometry}{\label{s_model}}

\subsection{Observational data}{\label{obs}}
We used imaging observations obtained with three large telescopes. This dataset contains 22 FORS2 frame series in 2013--2014 obtained by \citet{Boffin} with the high-resolution mode and a 0.126\arcsec/px pixel scale. The dates of  observations  and the average full width at half-maximum (FWHM) are given in Table \ref{fx}. The 2015 observations were obtained when the relative binary separation was $\sim$0.7\arcsec, which is too small for very precise astrometry, and we therefore did not use these data.

\begin{table}[tbh]
\caption{ Dates of FORS2 observations, average  FWHM $ {\varepsilon} $ (in arcseconds), and DCR-related functions used in Eq. (\ref{eq:model}). }
{\tiny
{\centering
\begin{tabular}{@{}ccccccc@{}}
\hline
\hline
epoch & Year           & $ {\varepsilon}$& $f_{1,x}$   & $f_{2,x}$   &$f_{1,y}$   &$f_{2,y}$ \rule{0pt}{11pt}\\
\hline
  1 &2013.28541 &0.62 & -0.6030 & -0.7187&  -0.4672&  -0.5568  \rule{0pt}{11pt} \\
  2 &2013.29928 &0.67 & -0.0476 & -0.0585&  -0.5484&  -0.6746 \\
  3 &2013.31565 &0.55 & -0.0724 & -0.0873&  -0.5477&  -0.6602 \\
  4 &2013.34301 &0.87 &  0.0541 &  0.0656&  -0.5482&  -0.6644 \\
  5 &2013.35959 &0.79 &  0.5369 &  0.6555&  -0.4838&  -0.5907 \\
  6 &2013.38392 &0.83 & -0.0526 & -0.0656&  -0.5483&  -0.6835 \\
  7 &2013.39776 &0.64 &  0.4096 &  0.5067&  -0.5108&  -0.6319 \\
  8 &2013.41408 &0.69 &  0.2562 &  0.3176&  -0.5339&  -0.6620 \\
  9 &2013.42505 &0.71 &  0.3719 &  0.4497&  -0.5175&  -0.6257 \\
 10 &2013.43866 &0.68 &  0.2840 &  0.3391&  -0.5305&  -0.6334 \\
 11 &2013.44969 &0.73 &  0.5133 &  0.6454&  -0.4894&  -0.6152 \\
 12 &2013.45784 &0.69 &  0.4349 &  0.5260&  -0.5060&  -0.6119 \\
 13 &2013.47425 &0.90 &  0.4995 &  0.6075&  -0.4925&  -0.5989 \\
 14 &2014.09650 &0.72 & -0.1541 & -0.1876&  -0.5436&  -0.6619 \\
 15 &2014.12097 &0.58 & -0.3863 & -0.4714&  -0.5151&  -0.6285 \\
 16 &2014.18656 &0.72 & -0.2443 & -0.2937&  -0.5354&  -0.6436 \\
 17 &2014.21130 &0.80 &  0.1625 &  0.1985&  -0.5430&  -0.6632 \\
 18 &2014.24126 &0.68 & -0.0222 & -0.0272&  -0.5489&  -0.6737 \\
 19 &2014.27135 &0.68 &  0.1208 &  0.1465&  -0.5457&  -0.6618 \\
 20 &2014.31484 &0.64 & -0.3605 & -0.4392&  -0.5194&  -0.6327 \\
 21 &2014.34219 &0.88 & -0.2521 & -0.3028&  -0.5345&  -0.6419 \\
 22 &2014.37776 &0.80 & -0.0450 & -0.0548&  -0.5486&  -0.6681 \\
\hline
\end{tabular}   
\label{fx}          
}}
 \end{table}

We used the 36 HST measurements published in \citet{Bedin} that were obtained with the Wide Field Camera 3 in 2014--2016, which increases the total time span of high-precision observations to 3.5 years with approximately even sampling. This dataset is supplemented with six series of images obtained by \citet{Ammons} in 2014--2015 with Gemini/GeMS. Because GeMS uses adaptive optics, it  produces well-separated and therefore well-measured images of the components.  In addition, these infrared observations have low sensitivity to differential chromatic refraction (DCR) effects.  Some GeMS images were taken close in time to FORS2 epochs, which is useful to independently control the quality of calibrations in the FORS2 astrometry. The ESO archive contains some more archived images, for example,  photographic R-band images obtained with the Schmidt telescope in 1984.  We used these data,  reduced to ICRF   by \citet{Garcia}, to improve  the quality of our reduction. In total, we used 64 frame series  covering 3.5 years, and one more distant epoch of ESO-R in 1984.

Gaia DR2 includes two-parameter solutions, that is,\ positions only, for the two components of LUH 16. The identifiers are \object{Gaia DR2 5353626573562355584} for LUH 16 A and \object{Gaia DR2 5353626573555863424} for LUH 16 B. Because of the large astrometric uncertainties (2--47 mas) and the potential biases in the DR2 positions due to unaccounted orbital motion and blending, we did not use these data in our analysis.

\subsection{Layout of investigation}{\label{lay}}

Several studies \citep{Boffin, 2015MNRAS, Garcia} used the same FORS2 dataset (Table \ref{fx}) but arrived at discrepant astrometric results. This illustrates the difficulties of ground-based sub-milliarcsecond astrometry, which are related to DCR, biases in measuring positions of close binary components, and other effects analyzed below. 

Here we aim at  mitigating these effects. We begin with a basic astrometric reduction  applied to the FORS2 observations, which  mitigates atmospheric image motion, geometric field distortion, DCR, and other effects to a level of about 0.1~mas. We started from the measured positions of stars $x'$, $y'$ obtained in our previous study \citet{2015MNRAS}. The reduction was performed in a sequence of repeating steps, which is necessary to estimate the optimal radius of the reference field and remove systematic errors  \citep{palta1, vb10, palta2}.

The following analysis is specific to LUH 16, and we used FORS2, HST, and GeMS observations to apply a model that includes new calibrations of the FORS2 measurements.  Solutions were derived iteratively by splitting the full model into independent blocks, finding approximate solutions for model parameters within these blocks, and repeating computations until converging to the final solution.  This expanded model includes
\begin{itemize}
\item  five astrometric parameters    $x_{c}$,  ${y}_c$,  $\mu_x$, $\mu_y$,  $\varpi$ for the barycenter motion of the binary, 
\item two DCR parameters $\rho$, $d$, which affect the barycenter motion as seen through the atmosphere (FORS2 only)
\item seven orbital elements, 
\item the mass ratio $q$,
\item 12  parameters $k_0^x \ldots k_2^x $, $k_0^y \ldots k_2^y $, for each binary component and each annual period, to calibrate for the bias in positions caused by the seeing and flux variations (FORS2 only), 
\item offset $\varepsilon_0$ for the effective average seeing  (FORS2 only), 
\item two calibration parameters to remove the offsets between the equatorial positions of the binary barycenter of FORS2 and HST.
\end{itemize}

In Sect.\,\ref{basic}  we describe the transformation of the measured positions $x'$, $y'$ to raw  $x_{raw}$, $y_{raw}$ positions  of each LUH 16 component in the reference frame of FORS2, free of the optical distortions. The positional residuals $\Delta'$ of the least-squares fit derived at this step provide information on the relative displacement of CCD chips \citep{palta2},  the  residual   correlations in the CCD space (Sect.\,\ref{s_m_cart}), and on variations with seeing (Sect.\,\ref{s_cart}). Calibration of these effects produces `clean` astrometric positions $\bar{x}$, $\bar{y}$, that are still affected by DCR, however (Sect.\,\ref{s_cart}). Some of the above calibrations require knowledge of the orbital motion of the LUH 16 components on the sky, however, which was derived by iteration starting from a rough initial approximation. 

The FORS2 and HST/GeMS positions were obtained in different reference systems. Whereas the FORS2 reference frame is local and set by the field stars, the HST/GeMS positions are given in ICRF. Therefore, adopting ICRF as a common system for this investigation, we transformed $\bar{x}$, $\bar{y}$  to  positions $x_A$, $y_A$,  $x_B$, and $y_B$ for each A and B component in ICRF  (Sect.\,\ref{ICRF}). These positions,  combined with HST and GeMS data,  were used to derive the solution for the orbital and barycenter motion (Sect.\,\ref{s_m_orb}).

\subsection{Basic astrometric reductions}{\label{basic}}

To reduce the  FORS2 observations, we employed our usual method, which ensures an astrometric precision of about 0.1~mas for isolated sources \citep{palta1, vb10, palta2}. The measured  {photocenter} positions $ {x}'$ and $ {y}'$ for every star in the field (reference or target) imaged at  time $t$ are represented by the model
\begin{equation}
\begin{array}{@{}l}
\label{eq:model}
     x_{0}+\hat{x}_0+ {\Phi}_{ x}(x',y') + \hat{\mu}_x (t-t_0)+ \hat{\varpi} p_x -
       \hat{\rho} f_{1,x}-  \hat{d} f_{2,x}     =  {x}'     \\
     y_{0}+\hat{y}_0+ {\Phi}_{ y}(x',y') + \hat{\mu}_y t(t-t_0)+ \hat{\varpi} p_y +
       \hat{\rho} f_{1,y}+  \hat{d} f_{2,y}    =  {y}'  
\end{array}
,\end{equation}
which describes the displacement of the stellar photocenter in the three-dimensional space formed by the coordinate axes  $x$, $y$ oriented along RA, Dec, and time $t$.  Here  $x_{0}$, $y_{0} $  are the approximate { CCD reference  star positions,} which are fixed  and define the directions for the coordinate axes  $x$, $y$ of the reference frame  at  the adopted  reference time $t_0= 56500$~MJD (=J2013.56742)  close to the average epoch.  Eq. (\ref{eq:model}) is solved by the least-squares fit, using all available measurements of reference stars. This yields  seven  astrometric parameters for each star:  the  coordinate offsets $\hat x_0$ and  $\hat y_0$, the proper motion  $\hat{\mu}_x$  and  $\hat{\mu}_y$,   the relative parallax  $ \hat{ \varpi }$, the DCR index $\hat{\rho}$ describing atmospheric displacement of the stellar image depending on the star's color, and the reverse displacement $d$, produced by the longitudinal atmospheric dispersion compensator (LADC) \citep{Avila}.  Eq. (\ref{eq:model}) also contains  the parallax factors  $p_x$,  $p_y$ and the DCR related functions  $f_{1,x}=\tan z  \sin \gamma$,   $f_{1,y}=\tan z  \cos \gamma$,   $f_{2,x}=\tan z_{\rm L} \sin \gamma$, and  $f_{2,y}=\tan z_{\rm L} \cos \gamma$, where $z$ is the zenith distance,  $z_{\mathrm L}$ is  the LADC parameter, and $\gamma$ is an angle between direction to zenith and north. 

The key component of Eq. (\ref{eq:model}) is the function $\Phi(x',y')$, which is derived from reference stars only and models  the sum of atmospheric image motion and geometric distortion {variations} for each individual frame. It is a full polynomial  of order $k/2-1,$ where the model parameter  $k$ is an even integer between 4 and 16. Here we set $k=10$ as the value that yields the most stable results. Using solution for $\Phi(x',y')$, we determined raw  positions 
\begin{equation}
\label{eq:rf}
     x_{raw} = x' -  {\Phi}_{ x}(x',y'); \quad      y_{raw} = {y'} - {\Phi}_{ y}(x',y') 
\end{equation}
of each A and B component in the reference frame, which are the measured positions  free of geometric distortion {variations}. Then from Eq. (\ref{eq:model}) we derived all astrometric parameters of the target objects (coordinate offsets, proper motion, parallax, and DCR parameters).

The LUH 16 components perform a nonlinear motion in the sky that is due to the curvature of the orbit, and the positional residuals $x_{raw}$, $y_{raw}$ display the measurement of the orbital segment $\psi$ expressed as $\psi_x=BX+GY$, $\psi_y=AX+FY$ with the Thiele-Innes parameters $A$, $B$, $F$, and $G$. The measured segment  represents the residual of $\psi$ affected by the least-squares fit (Eq. \ref{eq:model}).
Because of the correlation between the parameters of the system Eq.(\ref{eq:model}) with $\psi$,  their estimates are significantly biased. To emphasize this feature, we assigned a 'hat' to the  parameters of Eq. (\ref{eq:model}), in contrast to the 'actual' corresponding parameters
introduced later. The effect is illustrated by the difference of 3.53~mas in parallax values derived for components A and B with this preliminary approach, which is larger than the measurement uncertainty of 0.07 mas and is unphysical.

\subsection{Correlations across CCD}{\label{s_m_cart}}
The raw positions $  x_{raw}$, $  y_{raw}$ were corrected for space-dependent systematic errors as described below.
The residuals $\Delta$ of the least-squares fit of Eq. (\ref{eq:model}) for  reference stars are nearly random values with a scatter that corresponds to a typical FORS2 astrometric precision of 0.1--0.2~mas. However, they quite often display a correlation pattern of different type across CCD space. These correlations are traced and used to calibrate the raw positions in Eq. (\ref{eq:rf}) of LUH 16 for similar systematic errors. In this way, we detected and removed the relative displacement between CCD chips 1 and 2, which can affect the positions by 1--2~mas. Another type of systematic errors is seen as an oscillating autocorrelation function across CCD space. Correlations occur as a natural consequence of the polynomial fit of any measurements limited in space and therefore are not specific for FORS2.  For stars without orbital motion, these errors were mitigated as described in \citep{palta2}, but for LUH 16, the effect is masked by the dominating signal $\psi,$ which should first be subtracted from $\Delta'$.
 As an initial approximation for  $\psi$, we used polynomials of time that smoothen the individual residuals $\Delta'$ within each annual period, allowing us to form the corrected residuals $\Delta'-\psi,$ which  were then investigated and calibrated for errors correlated across the CCD. Later on, we used the actual signal $\psi$  to refine the calibrations iteratively. 

\subsection{Correlation of  FORS2 positions with seeing}{\label{s_cart}}

The images of the components of LUH 16A and LUH 16B partly overlap because the separation between the sources {   (1.4\arcsec on 2013 and 1.0\arcsec  and 2014) was comparable to the FWHM  (Table \ref{fx}), which  was  about 0.69\arcsec\ in RA and 0.76\arcsec\ in Dec on average  for both periods.} We therefore expect that the measurements of  CCD positions can be biased depending on seeing. This effect should cause a correlation between the positional residuals  and the seeing for individual images.\\ 
We found that  this correlation is of low significance when determined for individual frame series, therefore we split the data into the years 2013 and 2014,  during which the separation between the A and B components changed only little. To fit the measured positional residuals $\Delta'_x-\psi_x$ and $\Delta'_y-\psi_y$ separately for 2013 and 2014, we used the linear expressions
\begin{equation}
\label{eq:k}
\begin{array}{@{}ll}
H_x(\varepsilon_x, I) =  k_0^x + k_1^x[\varepsilon_x -(\bar {\varepsilon}  +\varepsilon_0)] +k_2^x(I/I_0 -1)\\
H_y(\varepsilon_y, I) =  k_0^y + k_1^y[\varepsilon_y -(\bar {\varepsilon}+\varepsilon_0)] +k_2^y(I/I_0 -1)\\
\end{array}
 ,\end{equation}
where $\varepsilon_x$ and $\varepsilon_y$ stand for the FWHM along $x$ and $y$ axes,  $\bar {\varepsilon}$ is the average FWHM for the annual period,   $\varepsilon_0$ is an offset  discussed below, and $k_0$, $k_1$, and $k_2$ are the free model parameters. In addition to the linear dependence on $\varepsilon$, the model Eq.(\ref{eq:k}) includes a  small but statistically significant component proportional to the change in flux $I$ in the image  relative to its annual average flux $I_0$. 

\begin{table}[tbh]
\caption{Coefficients of Eq.(\ref{eq:k}) of the systematic bias in the photocenter position of components A and B.}
{
{\centering
\begin{tabular}{@{}c@{\quad}c@{\quad}c@{\quad}c@{\quad}c@{}}
\hline
\hline
           &    $k_1^x$               &       $k_2^x$              &       $k_1^y$             &       $k_2^y$  \rule{0pt}{11pt} \\
            & (mas/arcsec)       &    (mas)                      & (mas/arcsec)      & (mas)  \\
\hline
 year        &   \multicolumn{4}{c}{ component A  }  \rule{0pt}{11pt}\\  
\hline 
2013  &  1.58 $\pm$ 0.45 &  -0.19 $\pm$ 0.23  &    -0.26 $\pm$ 0.63 &    -0.74$\pm$ 0.31   \rule{0pt}{11pt}    \\
2014  &  4.35 $\pm$ 0.45 &  -0.32 $\pm$ 0.29  &    -1.90 $\pm$ 0.47 &    -0.65$\pm$ 0.27   \\
 \hline                                                           
 year       &   \multicolumn{4}{c}{ component B  }  \rule{0pt}{11pt}\\  
\hline
2013  &    -4.30 $\pm$ 0.51&  -0.71 $\pm$ 0.24 &    6.06 $\pm$ 0.69&     -0.48$\pm$ 0.31  \rule{0pt}{11pt}     \\
2014  &    -4.96 $\pm$ 0.56&  1.45 $\pm$ 0.34 &    5.41 $\pm$ 0.50&     -0.93$\pm$ 0.28  \\
\hline
\end{tabular}   
\label{table}          
}}
 \end{table}           

The coefficients  $k_0$, $k_1$, and $k_2$ determined by the least-squares fit of the residuals $\Delta'- \psi $ with the functions (\ref{eq:k}) were used to correct the measured values $x'$, $y'$, ${ x_{raw}}$, and ${y_{raw}}$,  thus obtaining the positions 
\begin{equation}
\label{eq:xbar}
\bar{x} = x_{raw} - H_x(\varepsilon_x, I) \quad \bar{y} = y_{raw} - H_y(\varepsilon_y, I),
 \end{equation}
which correspond to  the average observational conditions set by $\bar {\varepsilon}$ and $I_0$. In this way, we decreased the internal scatter of the positional residuals  within each frame series and within each annual period, { but did not change their average values in 2013 and 2014 significantly, thus the orbital parameters are hardly affected by this correction.}

However, this does not yet solve the problem because the measurements that are not biased by seeing should correspond not to  $\bar {\varepsilon} $ but to $\varepsilon=\bar {\varepsilon}+\varepsilon_0$ offset to  a better seeing  by some  unknown value $\varepsilon_0$.  The key parameters derived at this phase of the reduction are the linear coefficients $k_{1,x}$ and $k_{1,y}$ (for components A and B each) used later to determine $\varepsilon_0$ in Sect.\,\ref{s_m_orb} as a free parameter of the relative orbital motion.

The derived coefficients of Eq.(\ref{eq:k}) except for the offsets $k_0$  are given in Table \ref{table}.  These coefficients were initially obtained with the residuals $\Delta' -\psi$  using a polynomial approximation of $\psi$. Then we  iteratively improved this estimate by applying $\psi$ derived for the full solution of the orbital motion. We note that  the coefficients $k_1$ are larger for  component B, as expected, because it is fainter, and they are approximately equal for both annual periods because the distance between the components did not change much. The values have opposite signs for components A and B. Considering the sign of $k_1$, this means that for poor{  seeing,} the measured positions of  the components are offset in the direction away from the other component.  The amplitude of the corrections in the positions for the maximum variations in seeing of about $\pm 0.2$\arcsec\  can reach $\pm 1$~mas, which  significantly exceeds the measurement uncertainty. 

\begin{table*}[tbh]
\caption{FORS2 { Cartesian measured}  positions $x_A$, $y_A$ of the A and $x_B$, $y_B$ of the B component, corrected for variations of seeing and flux (assuming that $\varepsilon_0=0$), but  including the DCR displacements and{ the measurement uncertainties  $\sigma_A$, $\sigma_B$.}  Positions  $x^*_A$, $y^*_A$, $x^*_B$, $y^*_B$ are corrected for DCR,  {for the offsets $\Delta_x$, $\Delta_y$ given in Table \ref{param}, and for $\varepsilon_0=-0.192$\arcsec}. Positions are linked to ICRF relative to the reference position $\alpha_0=162.3107300 \degr$, $\delta_0=-53.3181500 \degr$  and expressed in mas.}
{\small
{\centering
\begin{tabular}{@{}c@{\;\;}c@{\;\;}c@{\;\;}c@{\;\;}c@{\;\;}c@{\;\;}c@{\;\;}|c@{\;\;}c@{\;\;}c@{\;\;}c@{\;\;}|cc@{}}
\hline
\hline
             &   \multicolumn{6}{c}{ { Cartesian measured}  positions Eq.(\ref{eq:astr_ICRF})  }& \multicolumn  {4}{|c}{ { Cartesian corrected}  positions Eq.(\ref{eq:xycorr}) } &\multicolumn  {2}{|c}{ Relative positions  } \rule{0pt}{11pt}\\  
\hline
epoch&  $x_A$      &  $y_A $   &$\sigma_A$& $x_B$      &     $y_B $ &  $\sigma_B$  &     $x^*_{A} $ &  $y^*_{A}$ &     $x^*_{B} $ &     $y^*_{B} $   &  $x^*_{B}-x^*_{A}  $ &  $y^*_{B}-y^*_{A}  $  \rule{0pt}{11pt}\\
 \hline
 1 &    948.159   &   -470.565    &   0.168 &    -96.714   &   533.782   &     0.186 &   961.003     & -480.931     &   -80.846     &   520.734 &   -1041.850  &  1001.664     \rule{0pt}{11pt}  \\
 2 &    892.440   &   -427.278    &   0.112 &   -148.442   &   573.002   &     0.146 &   893.355     & -440.483     &  -146.242     &   556.831 &   -1039.597  &   997.314     \\
 3 &    815.484   &   -379.241    &   0.129 &   -222.726   &   616.691   &     0.161 &   816.892     & -391.732     &  -219.955     &   601.255 &   -1036.847  &   992.987     \\
 4 &    700.253   &   -294.262    &   0.135 &   -333.117   &   693.398   &     0.171 &   698.817     & -306.945     &  -333.591     &   677.763 &   -1032.407  &   984.708     \\
 5 &    645.418   &   -241.530    &   0.165 &   -383.083   &   741.177   &     0.198 &   632.783     & -252.979     &  -396.289     &   726.973 &   -1029.073  &   979.952     \\
 6 &    543.313   &   -160.993    &   0.125 &   -482.652   &   815.556   &     0.163 &   544.391     & -174.662     &  -480.272     &   798.908 &   -1024.663  &   973.569     \\
 7 &    508.819   &   -117.213    &   0.155 &   -513.252   &   855.337   &     0.188 &   498.790     & -129.712     &  -523.457     &   839.986 &   -1022.247  &   969.698     \\
 8 &    455.684   &    -62.883    &   0.165 &   -563.765   &   903.971   &     0.199 &   449.296     &  -76.012     &  -569.840     &   887.914 &   -1019.136  &   963.926     \\
 9 &    426.926   &    -28.830    &   0.133 &   -589.991   &   934.574   &     0.168 &   418.335     &  -40.744     &  -598.623     &   919.806 &   -1016.958  &   960.550     \\
10 &    389.562   &     13.531    &   0.096 &   -624.969   &   973.095   &     0.138 &   383.173     &    1.737     &  -631.115     &   958.418 &   -1014.288  &   956.681     \\
11 &    370.447   &     47.682    &   0.118 &   -641.358   &  1003.940   &     0.152 &   357.397     &   35.189     &  -654.930     &   988.649 &   -1012.328  &   953.460     \\
12 &    349.497   &     71.526    &   0.119 &   -661.631   &  1025.470   &     0.155 &   339.482     &   59.864     &  -671.887     &  1010.992 &   -1011.369  &   951.127     \\
13 &    318.666   &    118.700    &   0.128 &   -688.965   &  1067.789   &     0.164 &   307.021     &  107.171     &  -701.060     &  1053.481 &   -1008.080  &   946.310     \\
14 &   -842.125   &   -244.181    &   0.096 &  -1719.878   &   508.402   &     0.133 &  -839.329     & -256.607     & -1714.799     &   492.844 &    -875.470  &   749.451     \\
15 &   -985.044   &   -242.271    &   0.076 &  -1858.255   &   501.228   &     0.119 &  -976.862     & -254.106     & -1847.053     &   486.348 &    -870.191  &   740.455     \\
16 &  -1360.813   &   -184.604    &   0.086 &  -2217.756   &   536.883   &     0.128 & -1356.139     & -196.414     & -2210.519     &   521.978 &    -854.380  &   718.392     \\
17 &  -1493.606   &   -142.295    &   0.093 &  -2344.012   &   570.673   &     0.134 & -1498.115     & -154.812     & -2347.243     &   555.022 &    -849.128  &   709.834     \\
18 &  -1664.867   &    -79.152    &   0.087 &  -2509.057   &   623.308   &     0.132 & -1665.090     &  -91.972     & -2507.415     &   607.332 &    -842.325  &   699.303     \\
19 &  -1821.348   &     -4.898    &   0.081 &  -2657.438   &   687.157   &     0.125 & -1824.836     &  -17.232     & -2659.514     &   671.687 &    -834.679  &   688.919     \\
20 &  -2045.156   &    119.769    &   0.078 &  -2872.045   &   796.582   &     0.119 & -2037.609     &  107.888     & -2861.560     &   781.644 &    -823.951  &   673.756     \\
21 &  -2161.897   &    205.624    &   0.099 &  -2982.425   &   873.325   &     0.136 & -2157.066     &  193.868     & -2975.007     &   858.478 &    -817.940  &   664.610     \\
22 &  -2294.995   &    322.713    &   0.086 &  -3105.068   &   977.118   &     0.122 & -2294.703     &  310.170     & -3102.839     &   961.426 &    -808.136  &   651.256     \\
\hline                                       
\end{tabular}                                
\label{xyabs}                                
}}                                            
 \end{table*}

\subsection{Transformation into ICRF}{\label{ICRF}}                                             
It is convenient to perform the following analysis in the observational plane, where we have derived   positions Eq. (\ref{eq:xbar})   related to the FORS2 reference frame.  The metric of this frame,  however, is deformed   by the optical distortion and other effects that are significant {even at the small separation between the LUH 16 components.}  Therefore we transformed these positions into an absolute system that is now well  represented  by the Gaia source catalogs of DR1 \citep{Lindegren:2016aa} or DR2 \citep{DR2}.  As a suitable epoch  for the transformation we adopted  the  epoch $T_{\mathrm {DR1}}$=J2015.0. The Gaia DR1 catalog was also used  by \citet{Bedin} and \citet{Garcia} as reference  to  measure the positions of LUH 16. In this way, we obtained all positions  in the ICRF  at the epoch  $T_{\mathrm {DR1}}$. 

\subsubsection{ {Transformation  to the DR1 epoch}}{\label{ep_dr1}}
As explained in \citet{gaia}, the transformation was made only with stars imaged on the upper chip because the bottom CCD chip2 of FORS2 is rotated  and shifted relative to the upper chip1,  and the  use of both chips  leads to  large residuals of about 50~mas.  
We tangent-plane projected the Gaia equatorial coordinates to the CCD plane{ at the point $X_0= 1025.739$, $Y_0= 109.640$~px  near the position of LUH 16 in chip1 at the average time of observations, which corresponds to  $\alpha_0= 162.31073\degr$, $\delta_0=-53.31815\degr$,} and compared the projected Gaia DR1 and DR2 source positions  with those in the FORS2 reference system. Positions of DR2 were computed at the epoch $T_{\mathrm {DR1}}$ using the DR2 proper motions. To bridge the epoch difference  between the reference  epoch  $t_0$ of FORS2  and the comparison epoch $T_{\mathrm {DR1}}$, we corrected the FORS2 positions with proper motions $\hat{\mu}_x$, $\hat{\mu}_y$ of  Eq.(\ref{eq:model}) known with about twice better internally consistent precision as compared to DR2 (but  systematically offset from  the absolute, see Sect.\,\,\ref{abs}).
                                             
The following least-squares fit of positional residuals  was made with  functions $F_n^{x}(\bar{ x}-X_0, \bar{ y}-Y_0)$ for RA, and $F_n^{y}(\bar{ x}-X_0, \bar{ y}-Y_0)$ for Dec,  which are a sum of   two-dimensional  polynomials of the maximum power  $n$ formed on coordinates   $\bar{ x}-X_0$ and $\bar{ y}-Y_0$ of reference stars only. We tested all polynomial orders, starting from a linear model with $n=1$.  The optimal order $n=4$ was found by using an F-test for the rejection of non-significant coefficients as described in \citep{gaia}.  Thus,   using  79 stars  identified as common between FORS2 and DR2 on the upper chip, we obtained the average rms of the residuals of  1.10~mas, which is approximately consistent with an average uncertainty  0.67~mas for Gaia DR2 stars and 0.38~mas for FORS2 stars at this epoch. A similar transformation was also made with DR1. In this case, with 73 common stars,  the average rms of the residuals DR1 - FORS2  increased to  2.03~mas, which is explained by the larger uncertainty of the DR1 positions. In the following, we proceed with the results based on Gaia DR2.

 { With above procedure, we derived coefficients of functions $F_n^x$, $F_n^y$. Then, for each epoch of observations, we computed rectangular positions of LUH 16 A and B  linked to the ICRF, which along RA are computed as }
\begin{equation}                             
\label{eq:astr_ICRF}                         
x_A=\bar{ x}_A +F_n^{x}(\bar{ x}_A-X_0, \bar{ y}_B-Y_0),  
x_B=\bar{x}_B +F_n^{x}(\bar{ x}_B-X_0, \bar{ y}_B-Y_0).
 \end{equation}                              
The expressions for $y_A$, $y_B$ along Dec are similar.  The transformation uncertainty, which depends on the covariation function of $F_n$, is 0.21~mas for the stars in the center of the reference field, but it degrades to 0.43~mas  for our targets  located at the edge of this field (we recall that this reference field is not centered on the target because  all stars in chip2 were rejected).  When the transformation to the common system is made with DR1, these uncertainties are about twice larger.  The computed positions Eq. (\ref{eq:astr_ICRF}) may differ, within the above uncertainty limits, by some offset $\Delta$ from the `real' positions in ICRF computed with error-free function $F_n$. Considering that the function $F_n$  changes slowly across the CCD, we can assume that this offset is approximately equal for components A and B and for all their measurements at different epochs. The offset $\Delta$, if exists, is canceled out in the distances between the components, but  affects the barycenter position by $\Delta$. Because its value can be non-negligible, we included $\Delta$ in the model Eq. (\ref{eq:c_orb}) for the motion of LUH 16.
                                             
For similar reasons, the HST positions of  \citet{Bedin}  can be offset from the ICRF by about 0.3~mas, thus we expect to find systematic discrepancies between FORS2 and HST measurements by $\sim 0.5$~mas.

In the left columns of Table \ref{xyabs} we give  the positions Eq.(\ref{eq:astr_ICRF}), linked to the ICRF, that are affected by the DCR  and  reduced to the average seeing and flux with Eq.(\ref{eq:k}) and the coefficients from Table \ref{table}, assuming that $\varepsilon_0=0$.

\subsubsection{Conversion of  proper motions and parallaxes to absolute motions and parallaxes}{\label{abs}}
By definition, the differential  proper motions $\hat{\mu}$ and parallaxes $\hat{\varpi}$ of FORS2 systematically differ from their absolute values  $\mu(\mathrm {abs})$ and $\varpi(\mathrm {abs})$ by some constant offsets $\Delta\mu$ and $\Delta\varpi$, thus $\mu(\mathrm {abs})= \hat{\mu}+ \Delta\mu$ and $\varpi (\mathrm {abs}) = \hat{\varpi} +\Delta\varpi$ \citep{Lazorenko2009,palta2}. While this is not important for deriving the  relative angular motion of the binary, these parameters of the barycenter derived in Sect.\,\ref{s_bar} need to be corrected to absolute values. In our previous studies of nearby sources \citep[e.g.,][]{palta1}, we derived statistical corrections for the FORS2 relative parallaxes $\varpi$ by comparing our data with a galaxy model \citep{Robin2003}. For LUH 16 we thus derived the correction term  $\Delta\varpi =0.28 \pm 0.01$~mas \citep{2015MNRAS}. 

\begin{figure}[tbh]
\includegraphics[width=\linewidth]{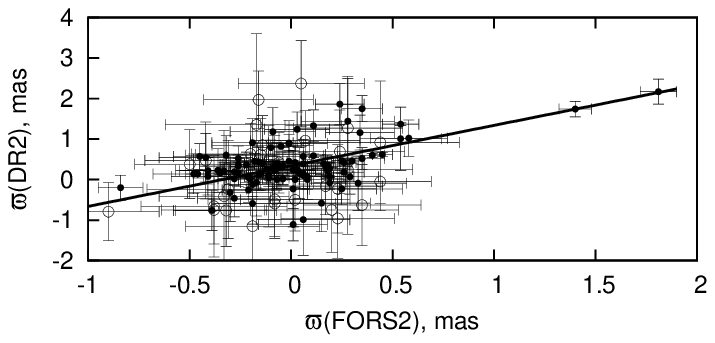}
\includegraphics[width=\linewidth]{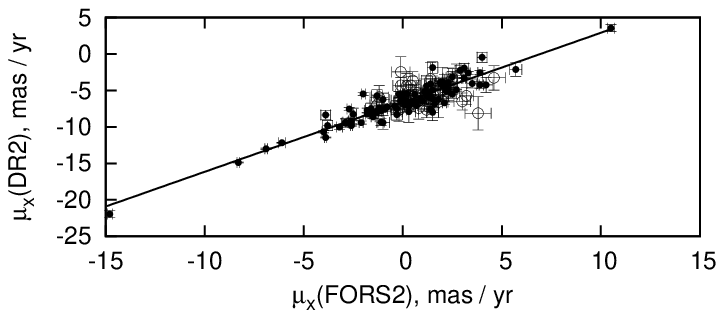}
\caption{Comparison between relative FORS2 and absolute DR2 parallaxes (upper panel) and proper motions in RA (lower panel) for brighter (filled circles; they were used to determine the correction) and fainter stars (open circles), with linear fit functions (solid lines). }
\label{comp}
\end{figure}

With the availability of Gaia DR2, it becomes possible to determine the parallax correction by directly comparing the FORS2 and DR2 parallaxes of field stars. {We used only those stars that define the system  of FORS2 astrometric parameters \citep[Sect. 3.3.1]{palta2}; they are typically brighter than $G=20$.  Because we used equal weights  $P_i$ in  \citet[Eq.2]{palta2} for these stars, the estimate of $\Delta\varpi $ is the simple arithmetic average offset. There were 65 stars in chip1 and  32 stars in chip2 that were also reduced to ICRF specifically for the purpose of deriving $\Delta\varpi $ and $\Delta\mu$.   This direct comparison  (Fig.\,\ref{comp}, upper panel) yields a parallax correction value of $\varpi_{\mathrm DR2}-\varpi_{\mathrm FORS2}=\Delta\varpi =0.342\pm 0.056$~mas, which within error bars agrees with the value reported by \citet{2015MNRAS}, thereby indirectly validating the model of \citep{Robin2003} and our previous statistical method of deriving $\Delta\varpi$ at a level of $\pm 0.1$~mas in this field.}

This assessment does not account for the DR2 parallax zero-point of $-29\,\mu$as discussed in \cite{DR2}, which was determined for quasars and may be variable in amplitude as a function of magnitude, position, and other parameters. Since Gaia parallaxes are therefore too small, the DR2 zero-point offset increases the value of the parallax correction we derive by the equivalent amount.

In a similar way, we compared the proper motions in RA  and obtained a linear dependence $\mu_x(\mathrm {DR2})=\Delta\mu_x + c_x\mu_x(\mathrm {FORS2})$  with an offset $\Delta\mu_x=-6.618 \pm 0.111$~{mas/yr} and a nearly unity coefficient  $c_x=0.953 \pm 0.037$~mas/yr  (Fig.\,\ref{comp}).  The RMS of the fit residuals of  $1.1$~mas/yr is small but exceeds the uncertainties of  0.18 and 0.47~mas/yr for proper motions of FORS2 and DR2, respectively. In Dec, using the similar  relation  $\mu_y(\mathrm {DR2})=\Delta\mu_y +c_y\mu_x(\mathrm {FORS2}) $, we derived   $\Delta\mu_x= 2.103 \pm 0.120$~{mas/yr},   $c_y=0.957 \pm 0.052$~mas/yr,  and a residual scatter of $1.2$~mas/yr. The calibration terms $\Delta\mu_x$, $\Delta\mu_y$, and $\Delta\varpi$ determined here were used later in Sect.\,\ref{s_bar} to convert the relative barycenter proper motion and parallax into the absolute values, neglecting potential Gaia DR2 proper motion zero-point offsets.

\section{Barycenter and orbital motion}{\label{s_m_orb}}
{ The measured positions $x_A$, $y_A$ $x_B$, $y_B$ derived with Eq.(\ref{eq:astr_ICRF}) contain directly measured target positions, all calibrations for the effects described above, and although it is not explicitly evident, contain a full information on all measurements of the reference stars.  The reference data were not discarded because they were integrated into the basic astrometric  model and then were extracted again at any exposure moment and any CCD pixel in a refined form, for instance, filtered from the atmospheric image motion.}

The measured positions from Eq.(\ref{eq:astr_ICRF}) were modeled by  the sum of the barycenter motion of the system and its orbital motion in RA,
\begin{equation}
\label{eq:c_orb}
\begin{array}{@{}l@{}l}
  x_A = &   x_{c} +\mu_{x}(t-t_0)  +  \varpi p_{x}+\lambda( - \rho_A f_{1,x} - d_A f_{2,x} 
 -k^x_{1,A} \varepsilon_0) \\
            &  +BX+GY  +\Delta_x \\
 x_B = & x_{c} +\mu_{x}(t-t_0)  +  \varpi p_{y}+\lambda( - \rho_B f_{1,x} - d_B f_{2,x} -k^x_{1,B}\varepsilon_0) \\
          & -Bq^{-1} X - Gq^{-1} Y   +\Delta_x,  \\
\end{array}
 ,\end{equation}
and similar for the positions in Dec $y_A$, $y_B$. The model includes   coordinate offsets $x_{c}$, $y_{c}$, proper motion $\mu_{x}$$(=\mu_{\alpha}\cos \delta)$, $\mu_{y}$, relative parallax $\varpi $,  Thiele-Innes parameters $A$, $B$, $F$, and $G$, and the mass-ratio $q$. The DCR parameters $\rho_A$, $\rho_B$, $d_A$, $d_B$, and the offset $\varepsilon_0$ to the average FWHM from Eq.(\ref{eq:k}) are applicable to FORS2 only, which we flag by $\lambda=1$ for FORS2, and $\lambda=0$  in the other cases. The constants $\Delta_x$, $\Delta_y$ are defined  in the previous  Sect.\,\ref{ep_dr1} and are the systematic offsets to the positions $x_A$, $y_A$, $x_B$, $y_B$, which depend on the instrument. These equations  can therefore contain $\Delta_{x}(\mathrm {HST})$,  $\Delta_x(\mathrm {FORS2})$, $\Delta_y(\mathrm {HST}),$ or $\Delta_y(\mathrm {FORS2})$. Because the observational information for GeMS is insufficient, we assumed that $\Delta(\mathrm {GeMS})=0$.

From Eq. (\ref{eq:c_orb}) we can obtain an expression for the barycenter motion in the sky that is independent of the orbital parameters, and an expression for the orbital motion that is independent of the barycenter motion. For the orbital motion, we have
\begin{equation}
\label{eq:orb}
\begin{array}{@{}ll}
    x_B - x_A  =& \lambda[ -(\rho_B - \rho_A)  f_{1,x} -(d_B - d_A) f_{2,x} \\ 
& -(k^x_{1,B}-k^x_{1,A}) \varepsilon_0] -(1+q^{-1})(BX+GY ) \\
    y_B - y_A  = &\lambda[ (\rho_B - \rho_A)  f_{1,y} +(d_B - d_A) f_{2,y} \\ 
&-(k^y_{1,B}-k^y_{1,A})  \varepsilon_0 ] -(1+q^{-1})(AX+FY ),
\end{array}
 \end{equation}
and for the barycenter motion, we find
\begin{equation} 
\label{eq:c2}
\begin{array}{@{}l}
\frac{ x_A+qx_B  }{1+q} =   x_{c} +\mu_{x}(t-t_0)  +   \varpi p_{x}
  -  \lambda( {\rho}  f_{1,x} + {d}  f_{2,x} +\tilde{k}^x_1  \varepsilon_0)+  \Delta_x  \\
\frac{ y_A+qy_B  }{1+q} =   y_{c} +\mu_{y}(t-t_0)  +  \varpi p_{y} 
  + \lambda({\rho}  f_{1,y} + {d}  f_{2,y}  -\tilde{k}^y  \varepsilon_0) + \Delta_y,
\end{array}
 \end{equation}
where ${\rho}  =(q \rho_B+\rho_A)/(1+q)$ and ${d}  =(q d_B+d_A)/(1+q)$ are the effective  DCR parameters that apply to the barycenter, and $\tilde{k}^x_1= (q k^x_{1,B}+k^x_{1,A}) /(1+q)$, $\tilde{k}^y_1= (q k^y_{1,B}+k^y_{1,A}) /(1+q)$   are the effective values of $k_1^x$ and $k_1^y$ . { Four individual parameter values  $\rho_A$, $\rho_B$,  $d_A$, and $d_B$ are obtained by combining the effective parameters $\rho$, $d$ and their differences $\rho_B - \rho_A$, $d_B - d_A$  derived from  Eq.(\ref{eq:orb}).}

\subsection{Deriving  orbital parameters}{\label{s_orb}}
\subsubsection{Deriving  $\varepsilon_0$}
\begin{figure}[tbh]
\centering
\includegraphics[width=\linewidth]{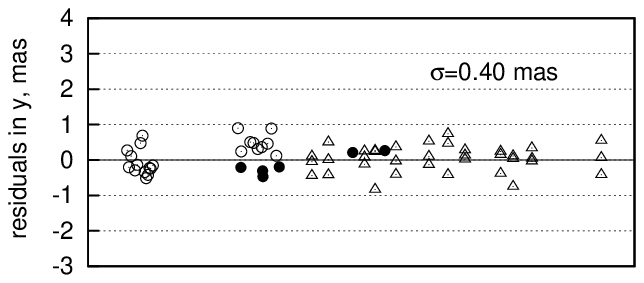}

\includegraphics[width=\linewidth]{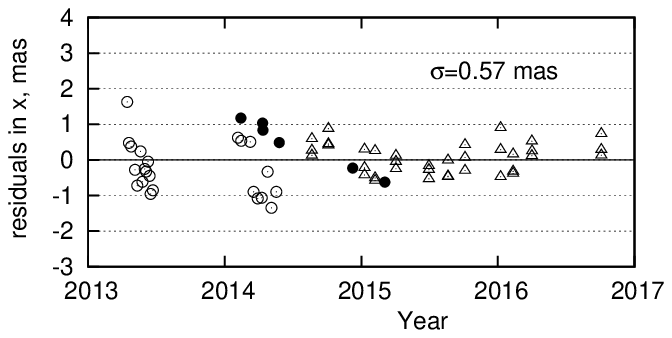}
\caption{Residuals in RA (lower panel) and Dec (upper panel) between the FORS2 (open circles), HST (triangles), and GeMS (filled circles) for the relative position of the secondary component of LUH 16  with $\varepsilon_0 =0$ in  Eq.(\ref{eq:orb}).  }
\label{dif_rez_ini}
\end{figure}

\begin{figure}[tbh]
\begin{tabular}{c}
{\includegraphics[width=\linewidth]{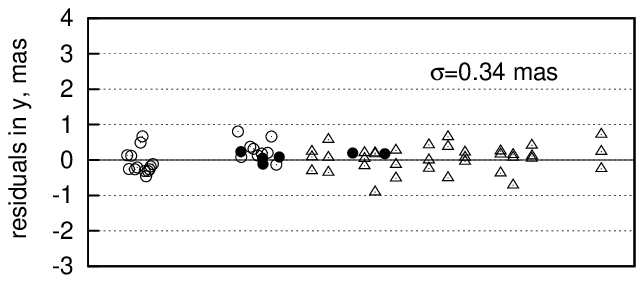}}\\
{\includegraphics[width=\linewidth]{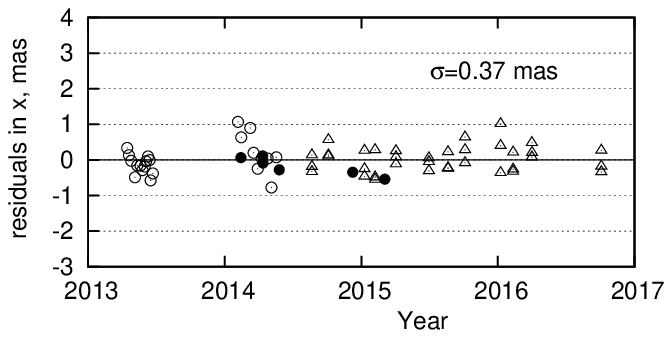}}
\end{tabular}
\caption{Same as Fig. \ref{dif_rez_ini}, but with $\varepsilon_0 =-0.192$\arcsec\ in  Eq.(\ref{eq:orb}).  }
\label{dif_rez_fin}
\end{figure}

{\label{eps0}}
The orbital parameters  were obtained by fitting FORS2, HST, GeMS, and ESO-R measurements with the model Eq. (\ref{eq:orb}). Because these equations are nonlinear, we initially used approximate values of the nonlinear parameters $e$, $T_0$, and $P$ spread over a wide range with a small increment and  computed the elliptical orbital coordinates $X$, $Y$ of component B relative to A at the observation time. Thus, for each set of $e$, $T_0$, and $P,$  we  linearized  Eq.(\ref{eq:orb})  in the parameters  $A$, $B$, $F$, and $G$, and $\rho$, $d$,  initially not including the term $\varepsilon_0$.  The least-squares  solution then yielded the rms  of the fit residuals, which indicates the likelihood  of the tested group of $e$, $T_0$, and $P$. As a final solution, we adopted the parameter set that yielded the best fit.  The positional residuals obtained in this way (Fig. \ref{dif_rez_ini}) show a clear systematic discrepancy between FORS2 and GeMS, and the rms of the residuals    $\sigma= 0.57$~mas is large. Therefore we complemented the model Eq.(\ref{eq:orb}) with the  term $\varepsilon_0$ and obtained $\varepsilon_0=-0.192 \pm 0.014$\arcsec. Thus we expect that at  seeing $ \bar{\varepsilon} +\varepsilon_0 = 0.50$\arcsec,\ the measurements are  not biased by the light from the nearby companion.  { The  new solution yielded  better  $\sigma= 0.357$~mas,  and the systematic offset between FORS2 and GeMS measurements obtained at comparable epochs was mitigated (Fig. \ref{dif_rez_fin}).}

\subsubsection{Initial and final solutions}{\label{rms}}
The barycenter and orbital motion  were initially computed with weights according to the measurement uncertainties $\sigma_{cat}$ associated with the data.  However, the measured rms of the positional residuals exceeds the  model expectation by a  factor of $K$ that is different for each data set and the barycenter or the orbital model. We therefore assumed that  $K \sigma_{cat}$ better represents  the real uncertainties,  and accordingly updated the weights  in Eqs.  (\ref{eq:orb}, \ref{eq:c2}). Then all computations in Sect.\,\ref{s_orb}, \ref{s_bar}  were repeated until convergence for $K$. Table\,\ref{exc} represents the summary  for the measured rms and $K$ values for each instrument and separately for the barycenter and orbital motion. This table shows that the residual rms is roughly equal for HST and FORS2, but the excess $K$ is  more significant in the barycenter motion for both space-  and ground-based astrometry, which can be caused by  some common problem, for example,  related to the conversion from CCD space into ICRF. In the following, we refer to the solutions with {$K=1$} for all data sets and with $K$ taken from Table\,\ref{exc} as the initial and final solution, respectively. {In this way,  the  sum of squares of the normalized residuals  $\chi^2$ decreased the high initial value of $\sim$305 to $\chi^2 = 130.00$, as expected for $2\times 65$ epoch measurements.}

\begin{table}[tbh]
\caption{Rms of positional residuals and the excess $K$ in the uncertainties  $\sigma_{cat}$ for each data set.}
\begin{tabular}{@{}c|@{\quad}c@{\;\;}c@{}|c@{\;\;}c@{\;\;}c@{\;\;}c@{}}
\hline
\hline
          &    \multicolumn{2}{c@{}|} {rms, (mas)}&    \multicolumn{4}{@{\quad}c} {$K$} \rule{0pt}{11pt} \\
Model                     & FORS2    &HST             & ESO-R&   FORS2 & GeMS & HST \\
 \hline
orbit      &   0.40    &   0.34\tablefootmark{*}  &0.57& 1.98 & 0.81 & 1.33       \rule{0pt}{11pt}\\
barycenter       &   0.24    &   0.39  & -          & 2.6           & -  &2.5  \\
\hline                                                                      
\end{tabular}   
\tablefoot{ \tablefoottext{*}{Average for HST  with GeMS.}}
\label{exc}          
\end{table}

\begin{table}[tbh]
\caption{Median  orbital parameters of LUH 16AB,  with 68.3\% confidence intervals for initial and final solutions  (with  $\sigma_{cat}$ and   $K\sigma_{cat}$ uncertainties  of the measurements, respectively),  and the corresponding data derived by \citet{Garcia}.  {The discrepancies in $i$ and $\Omega$ can be explained by different representations of the orbital parameters, as discussed in Sect.\,\ref{cmp}. The differences are removed by converting $i$ to $180\degr - i$ and $\Omega$ into $270\degr -\Omega$.}}
{\centering
\begin{tabular}{@{}l|@{\;\;}r@{\;\;}r@{\quad}r@{}}
\hline
\hline
parameter               & initial                          &final                     &{Garcia 2017}   \rule{0pt}{11pt}\\
 \hline
$a$ [mas]             & 1775$_{-41}^{+38}$              & 1784$_{-14}^{+13}$        & 1774$\pm$25            \rule{0pt}{11pt}\\
$a$ [AU]              & 3.539$_{-0.082}^{+0.076}$       & 3.557$_{-0.023}^{+0.026}$ & 3.54$\pm$0.05          \rule{0pt}{11pt} \\
$e$                   & 0.339$_{-0.014}^{+0.012}$       & 0.343$_{-0.005}^{+0.005}$ & 0.35 $\pm$0.02         \rule{0pt}{11pt}\\
$P$ [yr]              & 27.26$_{-1.30}^{+1.15}$         & 27.54$_{-0.43}^{+0.39}$   & 27.4 $\pm$0.25        \rule{0pt}{11pt}\\
$T_0$ [yr]            & 2017.81$_{-0.12}^{+0.15}$       & 2017.78$_{-0.05}^{+0.05}$ & 2017.8 $\pm$0.8       \rule{0pt}{11pt}\\
$i$ [deg]             & 100.23$_{-0.09}^{+0.08}$        & 100.26$_{-0.05}^{+0.05}$   &79.5 $\pm$0.25        \rule{0pt}{11pt}\\
$\omega$ [deg]        & 128.9$_{-3.7}^{+4.6}$           & 128.1$_{-1.5}^{+1.5}$      &130.4 $\pm$3.5         \rule{0pt}{11pt}\\
$\Omega$ [deg]        & 139.7$_{-0.12}^{+0.13}$         & 139.67$_{-0.05}^{+0.05}$  &130.12 $\pm$0.12      \rule{0pt}{11pt}\\
$M_{tot}$[$M_{Jup}$] &   62.52$_{-1.42}^{+1.86}$      & 62.06$_{-0.54}^{+0.57}$     &62.0 $\pm$1.9        \rule{0pt}{11pt}\\
$M_{A}$[$M_{Jup}$]   &   33.76$_{-0.77}^{+1.01}$      & 33.51$_{-0.29}^{+0.31}$     &34.2 $\pm$1.2        \rule{0pt}{11pt}\\
$M_{B}$[$M_{Jup}$]   &   28.76$_{-0.65}^{+0.86}$      & 28.55$_{-0.25}^{+0.26}$     &27.9 $\pm$1.0       \rule{0pt}{11pt}\\
\hline
\end{tabular}   
\label{orb_tbl}          
}
\end{table}

{Because the observed orbital segment is short, the range of solutions with nearly equal rms is relatively large. } We find that practically the same fit quality can be obtained when the parameters  $e$,  $P$, and  $T_0$ change by about $\pm 0.01$, $\pm 1$~yr, and $\pm 0.1$~yr, respectively.  We applied the approach of \citet{Lucy}, which is applicable to the fit  with a model that includes the subsets of a standard  nonlinear ($e$, $P$, $T_0$) and linear  ($A$, $B$, $F$, $G$) parameters\footnote{Alternatively, the orbital parameters and their confidence intervals can be derived using a Markov chain Monte Carlo method  \citep{mcmc} or permutation  \citep{2016perm, 2017perm}.}. We modified this method because for the particular case of  Eq.(\ref{eq:orb}), a linear group  has to include  $(\rho_B - \rho_A)$,  $(d_B - d_A)$, and $\varepsilon_0$, and thus contains  seven parameters. In contrast to the original algorithm of \citet{Lucy}, which takes into account the correlation between two pairs of parameters ($B$, $G$) and  ($A$,  $F$) only, we deal with the correlations between seven  parameters of a linear group as defined by their covariance matrix $\vec{ D}$. This difference was removed  by Jacobi   transformation \citep{Press}  of the  matrix $\vec{ D}$ into its diagonalized form  $\vec {U = V^{T} D V,}$ where $\vec {V}$ is a matrix whose columns are seven eigenvectors of $\vec{ D}$, and  the diagonal elements of $\vec{U} $ are the eigenvalues of $D$.  This allowed us  to closely follow the idea of  the \citet{Lucy} algorithm, using for transformation of the orthogonal variables $z$ \citep[Sect.\,A.4]{Lucy}   to  the linear subset of parameters \citep[Eq.A.11, Eq.A.15]{Lucy} an expression  $\vec {V  U^{1/2} z }$ where $\vec{U^{1/2}}$ is the diagonal matrix whose diagonal elements are $\sqrt{U}$. 

The resulting median values with 68.3\%  confidence intervals are given in Table\,\ref{orb_tbl}. The differences between the initial {($K=1$ for all data sets) and final ($K$ given in Table \ref{exc})} are smaller than the uncertainties of the latter.

{The} single-epoch ESO-R measurement from \cite{Garcia}, which was obtained approximately one orbital period before the main body of high-precision but short-term observations, {deserves particular attention}. When including that data point, the observations cover the full orbit and the uncertainties in the orbital parameters and masses improve by approximately a factor of three, in spite of its large nominal $\sigma_{cat}\approx 150$~mas (and final $K\sigma_{cat}\approx 85$~mas ) uncertainty. This agrees with the  finding of \citet{Garcia} that ESO-R data help to break the degeneracy between { their} model parameters. { In our case, we have sufficient data to avoid the degeneracy even without the ESO-R data point, but its incorporation improves the precision of our solution.}

\begin{figure}[tbh]
\includegraphics[width=\linewidth]{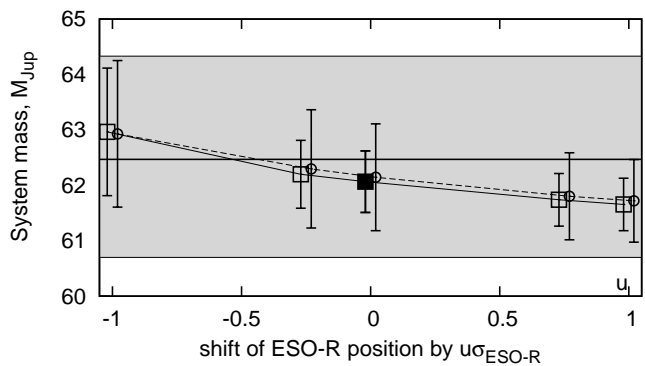}
\caption{ Median system mass with $1\,\sigma$ uncertainty ranges derived when applying an artificial shift to the ESO-R measurement. The shifts were applied in both coordinates by $u\sigma_{cat}$ toward ($u>0$) or away from ($u<0$) the model-predicted position, and the parameters   $K_\mathrm{ESO-R}=0.57$ (open square boxes) or $K_\mathrm{ESO-R}=1$  (open circles) were used. The adopted value of $M_{tot}$ given in Table\,\ref{orb_tbl} is shown as a filled square.  For comparison, the  $1\,\sigma$ uncertainty range (gray area) and the  median value (solid horizontal  line)  computed without the ESO-R measurements are shown.}
\label{ESO}
\end{figure}

In Table \ref{orb_tbl} we  give the total system mass  $M_{tot}=(a/\varpi_\mathrm{abs})^3/P^2$ and individual A and B masses computed with $q=0.8519$ and absolute $\varpi_\mathrm{abs}$ derived in the next section. The uncertainty of these dynamical masses is about 1\%, an approximately fourfold improvement compared to  \cite{Garcia}. Since the biases of the ESO-R measurement are difficult to quantify, { we performed several tests to asses potential systematic errors in the dynamical masses:   We varied the ESO-R measurement by \cite{Garcia} within the given uncertainty  shifting it in both coordinates by $u\sigma_{cat}$, towards ($u>0$) or away from ($u<0$) the model predicted position and repeated our analysis, see Fig.\,\ref{ESO}. We find that the precision of the orbital and mass parameters are hardly affected by the exact value of the ESO-R measurement, and the median values can vary within $\sigma_M$.  We therefore concluded that  our formal parameter uncertainties are correct and three to four times smaller than reported by \cite{Garcia}  if the ESO-R measurement is affected by random errors only. However, the accuracy of the individual masses depends on the systematic error/bias in the ESO-R measurement, which is unknown and may exceed the random errors (a case discussed in Sect.\, \ref{s_cart} in the context of FORS2). }This problem will likely disappear when the next orbital inflection point will have  been covered (in $\sim$2021), which will reveal the true bias in the ESO-R measurement.  

If the ESO-R measurement is not used, we obtain slightly higher masses of $M_{tot}=(62.6\pm 1.8) M_{Jup}$, $M_{A}=(33.8\pm 0.9) M_{Jup}$, and $M_{B}=(28.8\pm 0.8) M_{Jup}$, which are still within the confidence intervals for the final solution in Table\,\ref{orb_tbl}, but the uncertainties increased to 3\% { (Fig.\,\ref{ESO})}.

{  A minor source of bias in the masses may stem from the  seeing correction of the FORS2 measurements  (Sect.\,\ref{s_cart}), which partially   relies on the external HST and GeMS data.}

\subsubsection{Comparison to \citet{Bedin} and \citet{Garcia}}{\label{cmp}}

Except for eccentricity, our parameter values agree with \citet{Bedin} within the error bars, but our estimates have smaller confidence intervals because we have access to a longer time span ($\sim$3.5 instead of 2 years). Our estimates also agree with the median parameter values  obtained by \citet{Garcia} with the Markov chain Monte Calro (MCMC) analysis.  Because of a larger volume of observations and improved reduction of FORS2 data, our confidence intervals are 2--10 fold  narrower for most parameters, except  for the orbital period $P,$ for which the interval  27.1--27.6~yr given by \citet{Garcia} is smaller than the interval 27.1--27.9~yr derived by us. In contrast to \citet{Garcia}, we did not include radial velocity measurements in the fit, which may explain the slightly less well constrained period.

\begin{figure}[h!]
{\includegraphics[width=\linewidth]{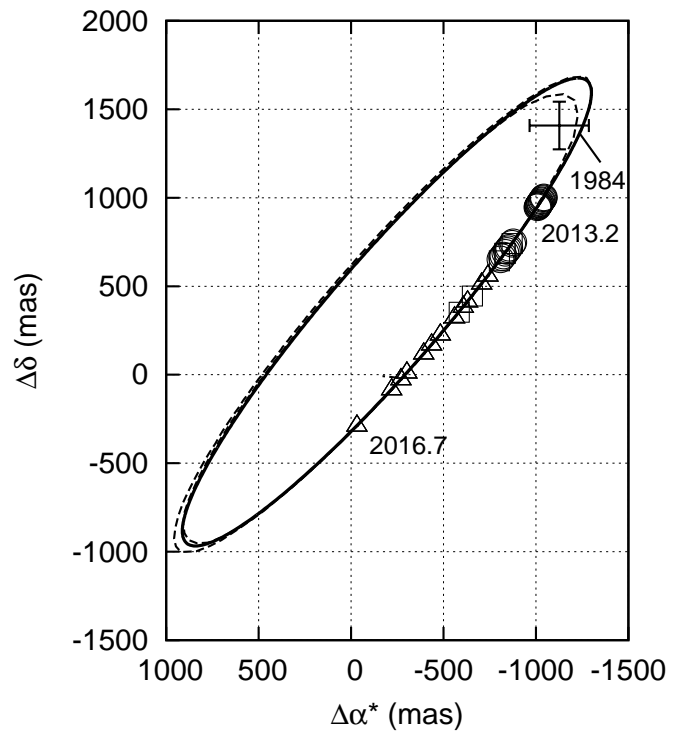}}
\caption{Motion of component B relative to A  for the direct data fit (solid line) for all solutions expected within the 68.3\% confidence interval (indicated by the space between the dashed lines), and the measured positions with FORS2 (open circles), HST (triangles),  GeMS (squares), and ESO-R {obtained in 1984} (the cross with error bars).}
\label{orb_demo}
\end{figure}

The inclination $i>90\degr$ that we obtained corresponds to the observed clockwise motion, whereas the value of $i<90\degr$  given by  \citet[Table 8]{Bedin} and \citet{Garcia} corresponds to a counterclockwise motion, {assuming identical parameterization. In particular, we define $\omega$ as the argument of periastron for the barycentric orbit of the primary}. The use of their orbital elements leads to equatorial positions with rearranged coordinate axes, that is,\ the computed RA is in fact Dec.  From  the definition of the Thiele-Innes parameters it follows that this inconsistency can be removed by changing  $i$ to $180\degr - i$ and $\Omega$ to $270\degr -\Omega$. For instance, the parameters  $i=79.21$ and $i=79.5\degr$,   and $\Omega=130.3$ and  $\Omega=130.12\degr$ given by \citet[Table 8]{Bedin} and \citet{Garcia}, respectively, are converted into $i=100.79$,  $i=100.5\degr$,   and $\Omega=139.7$,  $\Omega=139.88\degr$ which matches our estimates well. The relative  motion in the binary system is shown in Fig.\ref{orb_demo}, which essentially agrees with \citet{Garcia} also in terms of the families of allowed orbital configurations.

\begin{table}
\caption{ Radial relative velocity $V_A - V_B$ measured by CRIRES \citep{Garcia}  and the computed minimum/maximum values for  our orbital solutions within the 68.3\% confidence interval. }
\begin{tabular}{ccc}
\hline
\hline
Date      & $V_A - V_B$ (km/s)  & min / max value (km/s)\\
 \hline
2013.342  & 2.74$\pm$0.2    &  2.596  /  2.639     \rule{0pt}{11pt}\\
2014.333  & 1.94$\pm$0.2    &  1.915   / 1.957     \\
2014.382  & 1.85$\pm$0.2    &  1.871   / 1.914  \\
\hline                                                                      
\end{tabular}   
\label{rv}          
\end{table}           
{ We also computed the radial velocity of component LUH16 A relative to B for all possible orbital solutions within the 68.3\% confidence interval and compared these estimates (Table \ref{rv}) with the corresponding CRIRES  measurements presented by \citet{Garcia}. These velocity measurements were not included in the fit because they do not increase the time span of the data set and have a small weight compared to the orbital astrometry. However, we confirmed that our solution is compatible and that the inclination is $0<i<180^{\degr}$. We found good agreement between our model-predicted minimum/maximum velocity and the data (Table \ref{rv}) within the uncertainty range of the measurements.}

\begin{figure}[tbh]
\includegraphics[width=\linewidth]{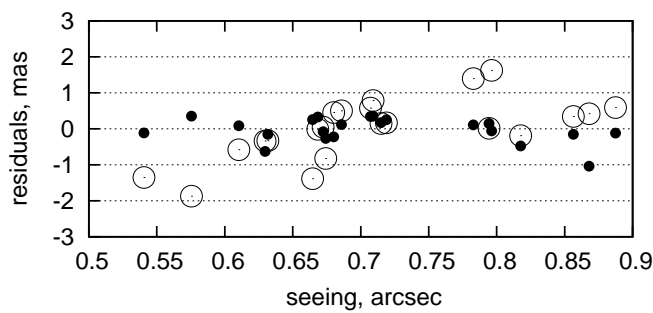}
\caption{Residuals along the  projected distance between components B and A   for  FORS2 measurements for our reduction (filled circles, $\sigma=0.35$~mas) and the reduction by \citet{Garcia} (open circles, $\sigma=1.2$~mas).  }
\label{g_f}
\end{figure}

{ We searched for a potential seeing dependence of the residuals $\Delta_{RA}$ in RA  and $\Delta_{Dec}$ in Dec  of the fit  Eq.(\ref{eq:orb}) that might be due to  errors in the photocenter measurements caused by overlapping images. Errors of this kind are expected to occur along the line connecting the components of LUH 16.  Therefore, projecting the residuals  $\Delta_{RA}$ and $\Delta_{Dec}$ onto this line, we formed  the equivalent residuals $\Delta_{||}$ , which we can correlate  with $\varepsilon$.  Because the binary system was oriented approximately at 45\degr relative to the coordinate axes, the projected  residuals are found simply as $\Delta_{||}= (\Delta_{RA}-\Delta_{Dec})/\sqrt{2}$, with the positive values corresponding to the displacement of LUH 16 B toward A.  Figure \ref{g_f} shows that there is no significant dependence of   $\Delta_{||}$ on seeing. 

For comparison,  we present  $\Delta_{||}$ computed based on HST, GeMS, and astrometric measurements  by \citet[Table 2]{Garcia}. With these data, we computed the orbital motion with Eq.(\ref{eq:c_orb}), neglecting the DCR terms  because these measurements are free from DCR, and obtained the  residuals  $\Delta_{RA}$ and $\Delta_{Dec}$ nearly equal to those shown by \citet[Fig. 10]{Garcia}. Then we formed the values $\Delta_{||}$ shown in  Fig. \ref{g_f} (open circles), which reveal a clear dependence on seeing. Their RMS of individual values is 1.2~mas, which significantly exceeds the value of 0.35~mas for our reduction. This comparison demonstrates the efficiency of our mitigation of seeing-dependent errors in the  epoch-average positions using statistics of residuals $\Delta' - \psi$ within individual frame series. }

\subsection{Deriving  the mass ratio and astrometric parameters}{\label{s_bar}}

\begin{figure}[tbh]
\begin{tabular}{c}
\includegraphics[width=\linewidth]{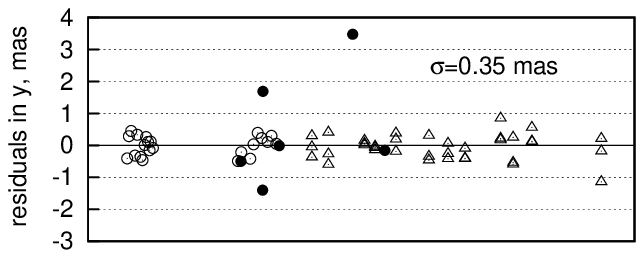}\\   
\includegraphics[width=\linewidth]{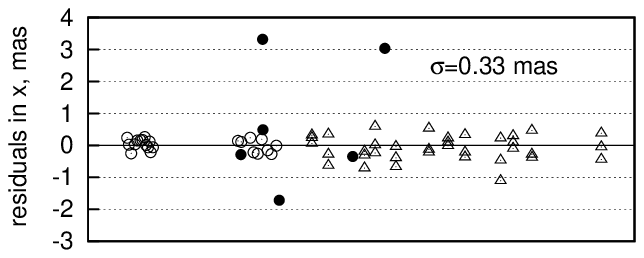}\\
\includegraphics[width=\linewidth]{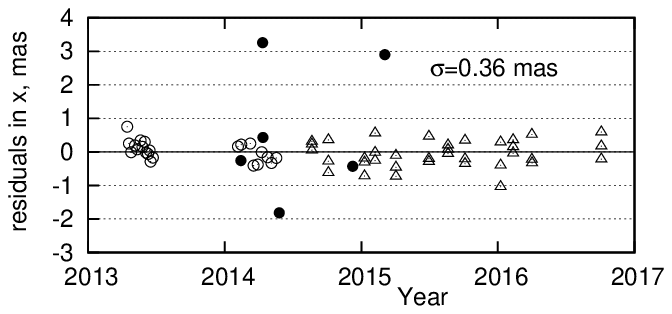}
\end{tabular}
\caption{Residuals  between the FORS2 (open circles), HST (triangles), and GeMS (filled circles) measured  barycenter position of LUH 16 and the model  of Eq. (\ref{eq:c2}).  The middle and upper panels are obtained assuming that  offsets $\Delta $ are different for each telescope, while    the lower panel shows the residuals in RA for the  standard model  of Eq. (\ref{eq:c2}) with no offset ($\Delta =0$). The $\sigma$ values refer to the average rms of residuals in RA and Dec for both HST and FORS2.}
\label{abs_rez}
\end{figure}

By fitting all HST, FORS2, and GeMS observations  with the model for the barycenter motion of Eq.(\ref{eq:c2}) in its simple form with $\Delta=0$ and $\varepsilon_0$ derived in Sect.\,\ref{s_orb}, we derived the first solution (left column of Table \ref{param}), yielding  $q=0.865$.  This value of $q$ gives  the smallest rms of 0.359~mas (0.352 and 0.366~mas along RA and Dec) for the residuals of all observed and model positions shown in  the lower panel of Fig.\ref{abs_rez} in RA.

\begin{table}[tbh]
\caption{Barycenter parameters   for { two versions of model Eq.(\ref{eq:c2}), without ($\Delta=0$), and with  separate offsets ($\Delta \neq 0$) in FORS2 and HST measurements, and the solution by \citet{Garcia}.  Zero-points $x_c$, $y_c$ are derived with reference to  $\alpha=162.3107300 \degr$, $\delta=-53.3181500 \degr$,         ICRF,  epoch J2015.0, and the zero time is $t_0 =56500$~MJD.  The precision $\sigma_{\mathrm{fit}}$  refers to the least-squares fit of the linear  Eq.-s (\ref{eq:c2}) with fixed $q,$ while  $\sigma_{\mathrm{stat}}$ takes into account  statistical variations of $q$ within $\pm 0.0024$.}}
{\tiny
{\centering
\begin{tabular}{@{}l@{\quad}r@{}l@{\;\;}r@{}l@{\;\; }l@{\;\;}r@{}l@{}}
\hline
parameter         & \multicolumn{2}{c} {no offset ($\Delta=0$)}&  \multicolumn{2}{c}{ separate offset} & &\multicolumn{2}{c} {Garcia 2017}    \rule{0pt}{11pt} \\
                      & \multicolumn{2}{c} {value  $\pm \sigma_{\mathrm{fit}}$}&  \multicolumn{2}{c}{value  $\pm \sigma_{\mathrm{fit}}$} &  $\sigma_{\mathrm{stat}}$ &\multicolumn{2}{c} {value  $\pm \sigma$}   \\
 \hline
$q$                &   0.8650    & $\pm$ 0.0024   &   0.8519  & $\pm$  0.0024  & \quad - & 0.82  & $\pm$  0.03        \rule{0pt}{11pt}\\
$x_{c}$   (mas)    &   24.371    & $\pm$ 0.066    &   28.338  &$\pm$    0.063  & 1.332    & -     &                    \\
$y_{c}$   (mas)    &  333.713    & $\pm$ 0.097    &   330.146 &$\pm$    0.104  & 1.190    & -     &                     \\
$\mu_x$   (mas/yr) &  -2759.709  & $\pm$ 0.045   &  -2760.884&$\pm$    0.077  & 0.384     &  -2762 &$\pm$ 2.5            \\
$\mu_y$  (mas/yr)     &  353.208 & $\pm$ 0.053   &  354.753  &$\pm$    0.075  & 0.481     &  358    &$\pm$ 3.5            \\
$\rho$     (mas)     &   37.582  & $\pm$ 6.64     &   39.129  &$\pm$     6.29  & 6.29     & 38.06  &$\pm$ 8               \\
$\rho_A$     (mas)   &  38.192   & $\pm$ 9.31     &  39.802   &$\pm$     7.99  &8.35      &  -    &                      \\
$\rho_B$     (mas)   &   36.877  & $\pm$ 11.58    &   38.339  &$\pm$     9.94  &10.40     &   -   &                        \\
$  d$     (mas)      &   -51.314 & $\pm$ 5.64   &   -52.199 & $\pm$    5.16  & 5.40       &   -48.73& $\pm$   7         \\
$  d_A$    (mas)   &   -50.897   & $\pm$ 7.66    &   -51.554 & $\pm$    6.55  & 6.80      &  -    &                      \\
$  d_B$    (mas)   &  -51.796    & $\pm$ 9.53     &   -52.956 & $\pm$    8.15  &  8.51    & -     &                      \\
$\varpi  $(mas)    &  501.165    & $\pm$ 0.055    &  501.215  &$\pm$    0.053  & 0.061    & 501.01 &$\pm$  0.51          \\
$\Delta_x$ (mas)    &   0        &               &  -0.095   &$\pm$    0.081  & 0.169     &  -    &                      \\
$\Delta_y$ (mas)    &   0        &               &  +0.303   &$\pm$    0.080  & 0.197     &  -    &                      \\
\hline                                                                      
rms        (mas)    &  0.359     &               &    \multicolumn{2}{c}{ 0.340} &  \quad  - &&   \rule{0pt}{11pt} \\
\hline
\end{tabular}   
\label{param}          
}}
\end{table}

{ To set the uncertainty limits for $q$, we used the Monte Carlo method, creating  random samples of  FORS2 and HST  measured positions, and processed them  with different values of $q$. For each sample,  we determined the best $q$ value as that which corresponds to the minimum rms of the positional residuals. The scatter of these  values around the average is characterized by the 1 $\sigma$ interval of 0.0024 and  corresponds to the statistical uncertainty of the $q$ value determination. With $q$ fixed to 0.865,  Eq.-s (\ref{eq:c2}) is solved by the least-squares fit as a linear system of equations,  and the uncertainty $\sigma_{\mathrm{fit}}$  of the parameters derived thus is given in  Table \ref{param}. 

Eq. (\ref{eq:c2}), however, is a nonlinear system if $q$ is considered as a free model parameter. The direct solution of this expanded model is same as the solution of the linear model with  $q=0.865$, but the  uncertainties of  parameters are quite different. They were estimated  by the above  Monte Carlo simulation when, together with $q$, for each random sample we also derived all parameters of the model of Eq.(\ref{eq:c2}).  Thus,  for  each parameter we obtained a set of solutions that corresponds to the expected statistical variations of $q$. Thus computed 1 $\sigma$  deviations $\sigma_{\mathrm{stat}}$ for each parameter are given in Table \ref{param}. Unlike the linear model, some parameters of the expanded model (e.g.,  $x_c$ and $\mu_x$) are highly correlated, which is indicated  by the inequality $\sigma_{\mathrm{stat}} \gg \sigma_{\mathrm{fit}}$. In parameter space, the values of these parameters are concentrated along a line with a small scatter corresponding to $\sigma_{\mathrm{fit}}$.}

The residuals for GeMS shown in  Fig. \ref{abs_rez} are qualitatively equal to those obtained by \citet[ Fig.12]{Garcia}. In comparison to  FORS2 and HST residuals, they are  more scattered, possibly because of insufficiently precise transformation into ICRF. Because of a large scatter $\sigma$,  the barycenter parameters of LUH 16 and the RMS of the residuals were computed  with zero weights for the GeMS data. However, we expect that the zero-point errors do not affect the distances between components A and B, and the scatter   in the  distances  measured with GeMS  and shown in Fig.\ref{dif_rez_fin}  are indeed much smaller than those in Fig.\ref{abs_rez}.  Indirectly, these features support our assumption on the origin of $\Delta$ in Sect.\,\ref{ICRF}, and that it is instrument dependent  and should be included in the model of Eq.(\ref{eq:c_orb}). 

The large uncertainties of the chromatic parameters $\rho$ and $d$  in Table\,\ref{param} are due to the large and expected correlation of about -0.9996, which means that  the  DCR displacement correction by the LADC is modeled by Eq.\,\ref{eq:model}  with an uncertainty smaller than 0.1~mas.

The high precision in the $x_{c}$, $y_{c}$ offset  determination  ($\pm 0.066$ and $\pm 0.097$~mas), which is much better than the uncertainty of the   transformation Eq.(\ref{eq:astr_ICRF}) into the ICRF of about 0.5~mas,  motivated us to try out a model function with  different offsets $\Delta(\mathrm {HST})$ for HST and  $\Delta(\mathrm {FORS2})$ for FORS2. This produced the second solution with better rms of 0.340~mas achieved at $q=0.8519$ (Table \ref{param}). In this model we cannot distinguish the offsets,  however, and instead  obtained  $x_c + \Delta_{x}(\mathrm {HST})=28.433 \pm 0.162$,  $x_c+ \Delta_{x}(\mathrm {FORS2})=28.243 \pm 0.063$,  $y_c +\Delta_{y}(\mathrm {HST})=329.843 \pm 0.159$ and $y_c +\Delta_{y}(\mathrm {FORS2})=330.449 \pm 0.104$~mas. Each of these sums corresponds to  the zero-point of the barycenter in the FORS2 or HST system, but they are significantly, about 3--4~mas, different from those {derived} for the first solution. We verified that this is due to the difference in $q$  but not because  different offsets $\Delta$ were introduced.  

The difference between the FORS2 and HST  zero-points $\Delta_{x}(\mathrm {FORS2}) - \Delta_{x}(\mathrm {HST})= -0.190 $ and $\Delta_{y} (\mathrm {FORS2})- \Delta_{y}(\mathrm {HST}) = +0.606$~mas is comparable to the uncertainty in the zero-points of the transformation into the ICRF, giving a potential explanation for the discrepancy. To confirm this, we derived the corresponding values for a transformation (Sect.\,\ref{ICRF}) using Gaia DR1 and obtained significantly larger differences $\Delta_{x}(\mathrm {FORS2}) - \Delta_{x}(\mathrm {HST})= -0.816$ in RA and $\Delta_{y} (\mathrm {FORS2})- \Delta_{y}(\mathrm {HST}) = +1.199$~mas in Dec. This demonstrates the difference between Gaia DR1 and DR2 and supports our assumption on the origin of this discrepancy.

Fig. \ref{abs_rez} shows the difference between the two solutions: with use of  common $x_{c}$,  $y_{c}$  (bottom panel) and offsets different for each data set (middle panel), which we illustrate here in RA.  Visually, the gain in overall RMS  is small ($\sigma$ decreased from 0.36 to 0.33~mas in RA), but for the subset of FORS2 residuals, the improvement is  more significant (0.31~mas versus 0.24~mas).  We applied {the F-test of additional parameters} to find that the improvement is statistically significant. With all FORS2 and HST measurements along RA and Dec, we derived  $F \approx 10$  with 2 and 114 degrees of freedom, which means that the simpler model is true with a probability $<10^{-4}$.

Finally, we converted the barycenter parameters obtained with separate offsets in Table\,\ref{param}  into the absolute system with the correction terms determined in Sect.\,\ref{abs}. We  obtained absolute proper motions and parallaxes of $\mu_x=  -2767.502\pm    0.147$~mas/yr,    $\mu_y=356.856  \pm    0.150$~mas/yr, and $\varpi_\mathrm{abs}= 501.557  \pm    0.082 $~mas.

\subsection{Corrected FORS2 positions}{\label{xy2}}
Now we can apply  corrections to the FORS2 measurements to remove the difference in offsets $\Delta$. Since we have no handle on the proportion with which the discrepancy occurs, the errors were assumed equal in magnitude but opposite in sign, that is, $\Delta(\mathrm {FORS2})= -\Delta(\mathrm {HST}) = \Delta$. The barycenter zero-point $x_c$, $y_c$ and the offsets $\Delta$ to the system of equatorial positions at the location of LUH 16 on the CCD obtained in this assumption are given in Table \ref{param}. The corrections $x(\mathrm {FORS2}) - \Delta_x$, $y(\mathrm {FORS2}) - \Delta_y$, $x(\mathrm {HST}) + \Delta_x$, $y(\mathrm {HST}) + \Delta_y$ reduce these two data sets of positions to a system anchored at the ICRF.  In Table \ref{xyabs} we give the positions 
   \begin{equation}
\label{eq:xycorr}
\begin{array}{@{}ll}
   x^*_A=  & x_A   + \rho_A  f_{1,x} + d_A  f_{2,x}  +k^x_{1,A} \varepsilon_0 -\Delta_x \\
  y^*_A=  & y_A   - \rho_A  f_{1,y} - d_A  f_{2,y}  +k^y_{1,A} \varepsilon_0 -\Delta_y \\ 
\end{array}
 \end{equation}
corrected also for DCR and  other calibrations, where the expressions for LUH 16 B are similar.

\section{Conclusion and discussion}{\label{conc}}

We presented an improved reduction of the FORS2 astrometric measurements of the LUH 16 binary and showed that they are consistent with the HST and GeMS positions in the literature. In our previous reduction  \citep{2015MNRAS}, the positions of  LUH 16 were linked to the ICRF using USNO-B,  and as noted by \citet{Bedin}, are inconsistent with the HST measurements at a level of 10--20~mas. This was due to a flawed transformation into the ICRF that used stars in both chips of FORS2. Before Gaia DR1, this effect could not be diagnosed and  lead to a significant bias in RA, Dec of about $\pm$50~mas \citep{gaia}. It had only a negligible effect on the relative positions of the LUH 16 components via the pixel scale and rotation of the coordinate axes, however.  The main reason for the inconsistency noted by \citet{Bedin} was the incorrect transformation into the ICRF with Eq.(\ref{eq:astr_ICRF}), where as the argument of functions $F_n$  we used the positions of  LUH 16 $\bar x$, $\bar y$ extrapolated to the USNO-B epoch J2000.0 instead of using the positions at the actual epochs.

Gaia DR2 provides the reference frame for determining absolute proper motions and parallaxes. However,  we argue that very large ground-based telescopes  provide us with competitive  differential astrometry for faint 16--21~mag stars. In the LUH 16 field analyzed in this paper, the relative proper motions and parallaxes were determined with FORS2 with an average precision of 0.18~mas/yr and  0.13~mas, respectively, for  16--21~mag stars in common with DR2.  In spite of the short one-year duration of FORS2 observations, this precision is about twice better than the corresponding values  0.48~mas/yr and  0.31~mas in the DR2 catalog. The good astrometric performance of FORS2 is due to the higher signal-to-noise ratio in the position estimation.

We refined the individual dynamical masses of LUH 16 to $33.5\pm 0.3\, M_{Jup}$  (component A) and $28.6\pm 0.3\,M_{Jup}$ (component B), which corresponds to a relative precision of $\sim$1\% and is three to four times more precise than the estimates of \citet{Garcia}. We found that{ a minor bias with $1\sigma$ of} the ESO-R resolved astrometry based on a 1984 photographic plate { leads to a change in the dynamical masses that does not exceed the random error} of our determination. The { exact characterization of any residual bias in mass} will be resolved in the future when the high-precision astrometry will cover a larger portion of the orbit.

Because of the importance of LUH 16 as an extremely well-studied system of nearby brown dwarfs, the refinement of the parallax, orbital, and dynamical mass parameters will continue, with the purpose of increasing the knowledge of the physical parameters of the system. We expect that the astrometric data set presented here will contribute significantly to this process.

\begin{acknowledgements}
This work has made use of data from the ESA space mission \emph{Gaia} (\url{http://www.cosmos.esa.int/gaia}), processed by the \emph{Gaia} Data Processing and Analysis Consortium (DPAC, \url{http://www.cosmos.esa.int/web/gaia/dpac/consortium}). Funding for the DPAC has been provided by national institutions, in particular the institutions participating in the \emph{Gaia} Multilateral Agreement.
\end{acknowledgements}

\bibliographystyle{aa}
\bibliography{luhm}

\end{document}